\documentclass[prd,final,twoside,onecolumn,preprintnumbers]{revtex4}

\newcommand{\dvec}[1]{\stackrel{\Rightarrow}{#1}\hbox to -0.35em{}}

\newcommand{\tr}{\:\mathrm{tr}\:}

\renewcommand{\d}{\partial}

\newcommand{\La}{\mathcal{L}}

\newcommand{\bmx}[1]{\left(\begin{array}{*{#1}{c}}}
\newcommand{\emx}{\end{array}\right)}
\newcommand{\bwmx}[1]{  
	\renewcommand{\arraystretch}{2} 
	\left(\begin{array}{*{#1}{c}} }
\newcommand{\ewmx}{\end{array}\right)}
\newcommand{\bwwmx}[1]{  
	\renewcommand{\arraystretch}{2.5} 
	\left(\begin{array}{*{#1}{c}} } 
\newcommand{\bmmx}[1]{  
	\renewcommand{\arraystretch}{1.5} 
	\left(\begin{array}{*{#1}{c}} }
\newcommand{\bdet}[1]{\renewcommand{\arraystretch}{2.5}
	\left|\begin{array}{*{#1}{c}}}
\newcommand{\edet}{\end{array}\right|\renewcommand{\arraystretch}{1}}
\newcommand{\beq}{\begin{equation}}
\newcommand{\eeq}{\end{equation}}
\newcommand{\bea}{\begin{eqnarray}}
\newcommand{\eea}{\end{eqnarray}}

\newcommand{\ds}{\displaystyle}

\newcommand{\rj}{\begin{rjesenje}}
\newcommand{\kr}{\end{rjesenje}}
\newcommand{\dd}{\,d}

\renewcommand{\vec}[1]{\ensuremath\mathbf{#1}}

\begin{document}

\preprint{MIT-CTP-3317}
\preprint{CCNY-HEP-3/02}
\preprint{BUHEP-02-31}

\title{Non-Abelian Fluid Dynamics in Lagrangian Formulation}
%
\author{B. Bistrovic}
\author{R. Jackiw}
\affiliation{
Center for Theoretical Physics, MIT, 
Cambridge, MA 02139-4307, USA}

\author{H. Li}
\author{V. P. Nair}
\affiliation{Physics Department, City College of the 
CUNY New York, NY 10031, USA}
\author{S.-Y. Pi}
\affiliation{Physics Department, Boston University, 
Boston, MA 02215, USA}

\begin{abstract}
\noindent 
Non-Abelian extensions of fluid dynamics, which can have
applications to the quark-gluon plasma, are given. These theories
are presented in a symplectic/Lagrangian formulation and involve
a fluid generalization of the Kirillov-Kostant form well known in
Lie group theory. In our simplest model the fluid flows with velocity
$\vec{v}$ and in  presence of non-Abelian chromoelectric/magnetic
$\vec{E}^a/\vec{B}^a$ fields, the fluid feels
a Lorentz force of the form $Q_a \vec{E}^a +(\vec{v}/c)\times Q_a
\vec{B}^a$, where $Q_a$ is a space-time local non-Abelian charge
satisfying a fluid Wong equation $[(D_t +\vec{v}\cdot\vec{D})Q]_a=0$
with gauge covariant derivatives.
\end{abstract}

\maketitle

\section{Introduction}
\par
\hspace{0cm}\indent When different species of particles compose
a fluid, it may be that the collective variables describing
the fluid (density $\rho$, velocity $\vec{v}$ in an Eulerian
formulation), and the dynamical equations that govern them,
can reflect the compositional variety. In this paper we develop
models for fluids where the variety of constituents arises from
an internal symmetry group. We envision applying our theory to
high densities of non-Abelian quarks, with or without additional
interaction to a dynamical gauge field.
\par
High energy collisions of heavy nuclei can produce
of a plasma state of quarks and gluons. This new state of
matter has recently been of great interest both theoretically
and in experiments at the RHIC facility and at CERN.  Most of
the theoretical work in this area has been based on perturbative
Quantum Chromodynamics at high temperatures with hard thermal
loop resummations and other improvements \cite{ref-pisarski}.
This can be a good description at high temperatures and for plasma
states that are not too far from equilibrium. An alternative
approach, which may be more suitable for nondilute plasmas or
for situations far from equilibrium, would be to use a fluid
mechanical description.
\par
It is well known that the equations of fluid mechanics can be
derived from particle dynamics by taking suitable averages of
Boltzmann-type equations. Specifically for the  quark-gluon plasma,
some work along these lines was done many years ago using single
particle kinetic equations \cite{ref-heinz}.  These kinetic
equations form a hierarchy, the so-called BBGKY hierarchy,
involving higher and higher correlated $N$-particle distribution
functions. To be able to solve them, one needs to truncate the
hierarchy, very often at just the single particle distribution
function.  Therefore, the feasibility of solving these equations
limits the kinetic approach to dilute systems near equilibrium,
where the truncation can be justified.  We might therefore expect
that the regime of validity of equations derived within kinetic
theory is likewise limited.  However, fluid dynamical equations can
also be derived from very general principles, showing that they
have a much wider regime of validity, and, indeed in practice,
we apply them over such a wider range. This is the `universality'
of fluid dynamics.
\par
In the context of a non-Abelian plasma, in analogy with the
ordinary fluid mechanics, we may therefore ask for an \textit{a
priori} derivation of a non-Abelian fluid mechanics, which
incorporates the non-Abelian degrees of freedom, coupling to a
non-Abelian gauge field, \textit{etc.} This theory may be valid for
dense, nonperturbative and nondilute systems.  Further, a canonical
or symplectic formulation (at least in the conservative limit)
is important for quantization. At the same time, the analysis
based on the kinetic equations remains useful to us as a guide
for arriving at Lagrangian, canonical description.
\par
Our goal is to provide plausible equations that generalize to
the non-Abelian situation the continuity
\beq\label{eq11}
\d_t \rho(t,\vec{r}) + \nabla\cdot \left( \rho(t,\vec{r})
\vec{v}(t,\vec{r})\right)=0
\eeq 
and Euler equations
\beq\label{eq12}
\d_t \vec{v} (t,\vec{r}) + \vec{v}\cdot
	\nabla \vec{v}(t,\vec{r})=\vec{force}
\eeq
which govern conventional Eulerian perfect fluids.  In (\ref{eq12})
$\vec{force}$ denotes forces acting on a unit volume of the fluid.
The above four equations or their relativistic versions can be
equivalently presented as the four conservation laws for the energy
and momentum densities (either nonrelativistic or relativistic).
\par
In the non-Abelian generalizations one must first fix the group
transformation law for the collective variables and the covariance
properties of the equations, which now may number more than four,
and cannot be comprised solely in energy-momentum conservation. An
obvious additional equation is the (covariant) conservation
equation for the non-Abelian current, but the relation of this
current to the collective variables needs to be established.
\par
Our constructions begin by postulating collective variable
Lagrangians that are invariant against space-time translations
and internal symmetry transformations.  Thus energy-momentum
conservation and current conservation are assured by Noether's
theorem, while the Euler-Lagrange equations describe the
temporal evolution of the collective variables. In the Section
\ref{sec2} we review the Lagrangian as well as other aspects of
nonrelativistic and relativistic Abelian fluids. This is done
to motivate and contrast our non-Abelian constructions, which
we develop in Sections \ref{sec3} and \ref{sec4}. Derivations
are postponed to the Appendices; in the text only results are
presented and discussed.

\section{\label{sec2}Review of Abelian Fluids}

\hspace{0cm}\indent Equations (\ref{eq11}) and (\ref{eq12})
can be found as the Euler-Lagrange equations for the Lagrange
density \cite{ref-jackiw-1}
\beq\label{eq21}
\La= -j^\mu a_\mu +\frac{1}{2}\rho \vec{v}^2-V
\eeq
where our coordinates are $x^\mu=(c t,\vec{r})$. The various
quantities are defined as
\bea\label{eq21b}
j^\mu&=&(c \rho, \,\rho\vec{v})  \\
a_\mu&\equiv&\d_\mu\theta + \alpha \d_\mu \beta
\label{eq21c}
\eea
and $V$ is a $\rho$-dependent potential that gives rise to the
$\vec{force}$. (In the usual theory, $V$ depends only on $\rho$,
but we allow dependence on derivatives of $\rho$ as well.) The
canonical 1-form is determined by the first contribution to $\La$
while the dynamics is encoded in the $\frac{1}{2}\rho \vec{v}^2-
V$ term. Specifically, variation of $\vec{v}$ shows that $\vec{v}$
is given by the Clebsch parametrization \cite{ref-clebsch}
\beq\label{eq22}
\vec{v}=\nabla \theta + \alpha \nabla \beta\;.
\eeq
Variation of $\theta$ results in (\ref{eq11}), while variation
of the Clebsch potentials $(\alpha, \beta)$ produces subsidiary
equations
\beq
\label{eq23}
\begin{array}{rcl}
j^\mu \ds\d_\mu \alpha &=& 
	\rho\left(\ds\d_t \alpha 
	+ \vec{v}\cdot \nabla \alpha\right) =0\\
j^\mu \ds\d_\mu \beta &=& 
	\rho\left(\ds\d_t \beta 
	+ \vec{v}\cdot \nabla \beta\right) =0
\end{array}
\eeq
which are needed in the subsequent derivation of the Euler equation.
Finally, the ``Bernoulli'' equation emerges by varying $\rho$
\beq\label{eq24}
\d_t \theta + \alpha \d_t \beta +\frac{\vec{v}^2}{2}
=-\frac{\delta}{\delta \rho} \int\dd r\: V\;.
\eeq
($\frac{\delta}{\delta \rho} \int \dd r \:V $ is just the
Euler-Lagrange derivative of $V$.) The Euler equation (\ref{eq12})
then follows by taking the gradient of (\ref{eq24}) and using
(\ref{eq23}). The force on the right of (\ref{eq12}) is now seen
to be
\beq
\vec{force}= -\nabla \frac{\delta}{\delta \rho} \int \dd r \:V
\eeq
The energy $\mathcal{E}$ and momentum $\vec{\mathcal{P}}$ densities
carried by the fields are
\bea
\mathcal{E}&=&\frac{1}{2}\rho\vec{v}^2 +V \\
\vec{\mathcal{P}}&=& \rho\vec{v}\;.
\eea
These satisfy continuity equations as a consequence of (or
equivalent to) (\ref{eq11}) and (\ref{eq12}).
\par
Interaction with electromagnetic fields is accommodated
by including the Maxwell Lagrange density and adding the
electromagnetic vector potential to $a_\mu$, whereupon $j^\mu$
becomes the electromagnetic current. This gives rise to
the Lorentz force in the Euler equations, thereby producing
magnetohydrodynamics. Note that in this model the mass density
moves with the same velocity as the charge density as is seen by
inspecting $\vec{\mathcal{P}}$ and $\vec{j}$.
\par
An interesting realization of the above equations is provided by
M\"adelung's ``hydrodynamical'' rewriting of the Schr\"odinger
equation
\cite{ref-madelung}
\beq\label{eq27}
i\hbar\: \d_t\psi(t,\vec{r})
	=-\frac{\hbar^2}{2m}\nabla^2\psi(t,\vec{r})
\eeq
where the wave function is presented as
\beq\label{eq28}
\psi(t,\vec{r})=\sqrt{\rho(t,\vec{r})} 
	e^{i m \theta(t,\vec{r})/\hbar}\;.
\eeq
The imaginary part of (\ref{eq27}) reproduces the continuity
equation (\ref{eq11}), when $\nabla \theta$ is identified as
the velocity $\vec{v}$, with vanishing vorticity $\nabla\times
\vec{v}=0$. Also the quantum current $\frac{\hbar}{m}\mbox{Im}\,
\psi^{*} \nabla \psi$ become $\rho\vec{v}$. The real part of
(\ref{eq27}) gives the Bernoulli equation with $\alpha$ and $\beta$
set to zero and with a ``quantum'' force derived from
\beq\label{eq28b}
V=\frac{\hbar^2}{2m^2} \left(\nabla\sqrt{\rho}\right)^2
=\frac{\hbar^2}{8m^2} \frac{\left(\nabla\rho\right)^2}{\rho}
\;.
\eeq
The Euler equation (\ref{eq12}) follows by taking the gradient
of the Bernoulli equation.
\par
The Lagrange density for a relativistic fluid is chosen as
\cite{ref-jackiw-1}
\beq\label{eq210}
\La=-j^\mu a_\mu -f(n)\;,\qquad 
	n\equiv\sqrt{{j^\mu j_\mu}/{c^2}}\;.
\eeq
The canonical 1-form is as in the nonrelativistic case
(\ref{eq21b})-(\ref{eq21c}) and now $j^\mu$ is written in the
Eckart form \cite{ref-degroot}
\beq\label{eq210b}
j^\mu=n u^\mu\;, u^\mu u_\mu =c^2 \;,\qquad u^\mu
	=(c, \vec{v})\! \bigm/ \! \sqrt{1-{v^2}/{c^2}} \;,\qquad 
n=\rho \sqrt{1-{v^2}/{c^2}}
\eeq
$f(n)$ encodes the dynamics; for the free theory $f(n)=n\,c^2$.
Variations of $\theta$, $\alpha$ and $\beta$ still produce
equations (\ref{eq11}) and (\ref{eq23}) while variation of $j^\mu$
evaluates $a_\mu$
\beq\label{eq211}
a_\mu=-\frac{u_\mu}{c^2}f'(n)
\eeq
where dash denotes differentiation with respect to the argument
of the function. We call (\ref{eq211}) the relativistic Bernoulli
equation, which leads to the Euler equation with the following
steps. The curl of (\ref{eq211}) reads
\beq
\d_\mu a_\nu - \d_\nu a_\mu =
	\d_\nu\left(\frac{u_\mu}{c^2} f'(n)\right) 
	-\d_\mu\left(\frac{u_\nu}{c^2} f'(n)\right)
\eeq
Since according to (\ref{eq22}) the left side is equal to $\d_\mu
\alpha \d_\nu \beta - \d_\nu \alpha \d_\mu \beta$, it vanishes by
(\ref{eq23}) when projected on $u^\mu$. Thus there remains the
relativistic Euler equation
\beq\label{eq212}
\frac{u^\mu}{c^2} \d_\mu\left( u_\nu f'(n)\right) 
	- \frac{u^\mu}{c^2} \d_\nu\left( u_\mu f'(n)\right) 
= \frac{u^\mu}{c^2} \d_\mu\left( u_\nu f'(n)\right) 
	-\d_\nu f'(n)
=0\;.
\eeq
It is easy to show that in the nonrelativistic limit, with 
\bea\label{eq-aux1}
n &\approx& \rho -\frac{1}{2 c^2}\rho \vec{v}^2 \\
u^\mu &\approx& (c , \vec{v})
\eea
and 
\beq\label{eq-aux2}
f=n c^2 +V(n)\;,
\eeq
the Lagrange density (\ref{eq210}) passes to (\ref{eq21}) (apart
from an irrelevant term $-\rho c^2$) and the spatial component of
(\ref{eq212}) reproduces the Euler equation (\ref{eq12}).
\par
The energy-momentum tensor for (\ref{eq210}) reads, after $a_\mu$
is eliminated with (\ref{eq211}) \cite{ref-6}
\beq\label{eq213}
T^{\mu\nu}= -g^{\mu\nu} \left( n\,f'(n) -f(n) \right) 
	+ \frac{u^\mu u^\nu}{c^2}\, n\, f'(n)\;.
\eeq
Its divergence $\d_\mu T^{\mu\nu}$ can be expressed as
\beq\label{eq215}
\d_\mu T^{\mu\nu} = \d_\mu \left(n u^\mu\right) 
	\frac{u^\nu f'(n)}{c^2} 
	+ n\left[ \frac{u^\mu}{c^2} \d_\mu\left( u^\nu
		f'(n)\right) -\d^\nu f'(n)\right]
\eeq
The first term on the right vanishes by the virtue of the
continuity equation (\ref{eq11}) and the second vanishes by
the relativistic Euler equation (\ref{eq212}); or vice versa:
since the two terms in (\ref{eq215}) are linearly independent
(the first is parallel to $u^\nu$ and the second is orthogonal
to it), conservation of $T^{\mu\nu}$ implies (\ref{eq11}) and
(\ref{eq212}).  Relativistic magnetohydrodynamics is achieved,
as in the nonrelativistic case, by adding the electrodynamical
potential to $a^\mu$.
\par
Finally we record one fact, which we shall use below, about the
alternative, Lagrange, formulation of fluid mechanics. Here one
describes the fluid with co-moving coordinates $\vec{X}(t,\vec{x})$
where $\vec{x}$ is a (continuously) varying particle label. It
may be chosen to coincide with $\vec{X}$ at $t=0$. The relation
between Euler and Lagrange variables is the following. For the
density we have
\beq
\rho(t,\vec{r})=\int \dd x \: \delta\left(\vec{X}(t,\vec{x})-\vec{r}\right)
\label{eq216}
\eeq
which is normalized to unity at $t=0$. (The integral and the
$\delta$-function follow the dimensionality of space.) Evidently,
in the course of the $\vec{x}$ integral, the $\delta$-function
evaluates $\vec{x}$ at an expression $\vec{\chi}(t,\vec{x})$ such
that $\vec{X}(t,\vec{\chi}(t,\vec{r}))=\vec{r}$; i.e., 
$\chi$ is the inverse of $\vec{X}$. There is also a Jacobian. Thus
\beq\label{eq217}
\frac{1}{\rho}=\left. \det \frac{\d X^i}{\d x^j}
	\right|_{\vec{x}=\vec{\chi}}
\eeq
The Euler velocity is given by 
\beq
\vec{v}=\left. \d_t \vec{X}\right|_{\vec{x}=\vec{\chi}}
\eeq
or 
\beq\label{eq218}
\rho(t,\vec{r})\vec{v} (t,\vec{r}) 
	= \int \dd x \: \d_t \vec{X}(t,\vec{x}) 
	\delta\left(\vec{X}(t,\vec{x})-\vec{r}\right)\;.
\eeq
A simple calculation shows that the density $\rho$ and current
$\vec{j}=\rho\vec{v}$ defined by these equations satisfy the
continuity equation (\ref{eq11}). [The Euler equation follows by
differentiating (\ref{eq218}) with respect to time, positing a
force that determines $\d^2 \vec{X}/\d t^2$, and performing the
corresponding $\vec{x}$ integral; we shall not need this here.]

%
%
\section{\label{sec3}Non-Abelian Models Based on a Particle
Substratum}
\subsection{\label{sec3a}Non-Abelian current}
\hspace{0cm}\indent{}Before presenting a specific model, we give
a general analysis of the non-Abelian current $J^\mu_a=(c\rho_a,
\vec{J}_a)$.
\par
The conventional formula for the current of a single,
non-Abelian point particle, moving in a 4-dimensional space-time
$\{t,\vec{r}\}$, along a space-time path $X^\mu(\tau)$ ($\tau$
parametrizes the path) is
\beq
J^\mu_a(t,\vec{r}) = c \int \dd \tau \: \mathcal{Q}_a(\tau) 
	\frac{\dd X^\mu(\tau)}{\dd \tau} 
	\delta (X^0(\tau)-ct) \delta(\vec{X}(\tau)-\vec{r})\;.
\label{eq31}
\eeq
This is covariantly conserved 
\beq\label{eq32}
\d_\mu J^\mu_a + f_{abc}A_\mu^b J^\mu_c\equiv  (D_\mu J^\mu)_a=0
\eeq
proved that $\mathcal{Q}_a$ satisfies the Wong equation
\cite{ref-wong}
\beq\label{eq33}
\frac{\dd \mathcal{Q}_a(\tau)}{\dd \tau} 
	+ f_{abc} \frac{\dd}{\dd \tau} X^\mu(\tau)
	\: A_\mu^b(X(\tau)) \mathcal{Q}_c(\tau)=0
\eeq
For several particles $\mathcal{Q}_a$ and $X^\mu$ are indexed by a
discrete particle label $n$, which is summed in the definition of
the current $J^\mu_a$. In a continuum limit $n\to \vec{x}$ we have
\beq\label{eq34}
J^\mu_a (t,\vec{r})= c\int\dd x \dd \tau\; 
	\mathcal{Q}_a(\tau,\vec{x}) 
	\frac{\d X^\mu(\tau,\vec{x})}{\d \tau} 
	\delta \left(X^0(\tau,\vec{x}) -ct\right) 
	\delta \left(\vec{X}(\tau,\vec{x}) -\vec{r}\right)
\eeq
\beq\label{eq35}
\frac{\d}{\d \tau}\mathcal{Q}_a(\tau,\vec{x}) + f_{abc} 
	\frac{\d}{\d \tau}X^\mu (\tau,\vec{x}) 
	A^b_\mu(X(\tau,\vec{x})) \mathcal{Q}_c(\tau,\vec{x})=0
\eeq
[Notice that, since we replace a sum over $n$ with an integral over
$\vec{x}$, $\mathcal{Q}_a(\tau,\vec{x})$ is now a charge density.]
The parametrization may be fixed at $X^0(\tau,\vec{x})=c\tau$
so that the equations read
\beq\label{eq36}
\begin{array}{rcl}
\rho_a(t,\vec{r}) &=& \displaystyle\int\dd x 
	\: \mathcal{Q}_a(t,\vec{x}) 
	\delta\left(\vec{X}(t,\vec{x})-\vec{r}\right) \\
\vec{J}_a(t,\vec{r}) &=& \displaystyle\int\dd x 
	\: \mathcal{Q}_a(t,\vec{x}) \d_t \vec{X}(t,\vec{x}) 
	\delta\left(\vec{X}(t,\vec{x})-\vec{r}\right)
\end{array}
\eeq
\beq\label{eq37}
\d_t \mathcal{Q}_a(t,\vec{x}) + f_{abc}\left[ 
	c A_0^b(t,\vec{X}(t,\vec{x})) + \d_t \vec{X}(t,\vec{x})\cdot
	\vec{A}^b(t,\vec{X}(t,\vec{x}))\right] 
	\mathcal{Q}_c(t,\vec{x})=0
\eeq
Observe that just as in the Abelian case discussed
above, the $\vec{x}$ integration evaluates $\vec{x}$
at $\vec{\chi}(t,\vec{r})$, the inverse of $\vec{X}$, and the
Jacobian factor $\left(\det \d X^i/\d x^j\right)^{-1}$ is just
the Abelian charge density $\rho$ [see (\ref{eq217}) ]. Thus
\beq\label{eq38}
\begin{array}{rcl}
\rho_a(t,\vec{r}) &=& Q_a(t,\vec{r}) \rho(t,\vec{r}) \\
\vec{J}_a(t,\vec{r}) &=& Q_a (t,\vec{r}) \rho(t,\vec{r}) 
	\vec{v}(t,\vec{r})
\end{array}
\eeq
or 
\beq
\label{eq38-1}
J^\mu_a(t,\vec{r})= Q_a(t,\vec{r}) j^\mu(t,\vec{r}) \;,
\eeq
where
\bea\label{eq38b}
Q_a(t,\vec{r})&=&\left. \mathcal{Q}_a(t,\vec{x})
	\right|_{\vec{x}=\vec{\chi}} \\
\label{eq38c}
\rho (t,\vec{r}) Q_a (t,\vec{r}) &=& 
	\int \dd x\: \mathcal{Q}_a (t,\vec{x})
	\delta\left(\vec{X}(t,\vec{x})-\vec{r}\right)
\eea
Moreover, differentiating (\ref{eq38c}) with respect to time and
using (\ref{eq11}) and (\ref{eq37}) results in an equation for
$\d Q^a/\d t$,
\beq\label{eq39a}
\d_t Q_a(t,\vec{r}) + \vec{v}(t,\vec{r})\cdot
	\nabla Q_a(t,\vec{r}) 
	= -f_{abc} \left(c A_0^b(t,\vec{r}) 
	+ \vec{v}(t,\vec{r})\cdot \vec{A}^b(t,\vec{r})
	\right) Q_c(t,\vec{r})
\eeq
which can also be written as
\beq\label{eq39b}
j^\mu (D_\mu Q)_a =0\;.
\eeq
This is analogous to the Abelian equations (\ref{eq23}). Equations
(\ref{eq39a}) and (\ref{eq39b}) can be understood from the fact
that the covariantly conserved current (\ref{eq32}) factorizes
according to (\ref{eq38-1}) into a group variable $Q_a$ and a
conserved Abelian current $j^\mu$.  Consistency of (\ref{eq11}),
(\ref{eq32}) and (\ref{eq38-1}) then enforces (\ref{eq39b}).
\par
We recognize that formulas (\ref{eq34}) and (\ref{eq36}) are the
non-Abelian version of the Lagrange variable -- Euler variable
correspondence [see (\ref{eq216}) and (\ref{eq218})]. Also,
(\ref{eq37}), (\ref{eq39a}) and (\ref{eq39b}) are the field
generalizations of the particle Wong equation (\ref{eq33}),
with (\ref{eq37}) being presented in the Lagrange formalism
and (\ref{eq39a}), (\ref{eq39b}) in the Euler formalism. The
decomposition of the non-Abelian current in (\ref{eq38-1}) is the
non-Abelian version of the Eckart decomposition (\ref{eq210b})
\cite{ref-heinz}. Indeed, (\ref{eq38-1}) may be further factored
as in (\ref{eq210b})
\beq\label{eq312}
J^\mu_a(t,\vec{r}) = Q_a(t,\vec{r})n(t,\vec{r}) u^\mu(t,\vec{r})\;,
\eeq
\par
In the remainder of Section \ref{sec3}, we are guided in
our construction of a dynamical model for non-Abelian fluid
mechanics and ``color'' hydrodynamics by the above properties
of the non-Abelian current, which follow from the very general
arguments, based on a particle picture for the substratum of
a fluid. In Section \ref{sec4} we present a different model,
based on a field theoretic fluid substratum.

%
%

\subsection{\label{sec3b}A Model for non-Abelian 
Color Hydrodynamics}

\hspace{0cm}\indent{}The model that we present is based on a group
with group elements $g$, and anti-Hermitian Lie algebra elements
with basis $T^a$.
\bea\label{eq313}
[T^a,T^b]=f^{abc} T^c\\
\tr (T^a T^b) =-\frac{1}{2} \delta^{ab}\;.
\eea
The Lagrange density for an Eulerian fluid built on such a group is
taken to be the following generalization of the Abelian expression
(\ref{eq210})
\beq\label{eq314}
\La= j^\mu 2 \tr \left[K g^{-1}D_\mu g \right] - f(n) + \La_{gauge}\;.
\eeq
Here $j^\mu$ is an Abelian vector field (current) which can also be
decomposed  as in (\ref{eq210b}) 
\beq\label{eq315}
j^\mu=(c\rho, \rho\vec{v})=n \,u^\mu\;,\qquad u^\mu u_\mu=c^2
\eeq
The covariant derivative 
\beq
D_\mu g=\d_\mu g +A_\mu g
\eeq
involves a dynamical non-Abelian gauge potential $A_\mu=A_\mu^a
T_a$ whose dynamics is provided by $\La_{gauge}$. $K$ is a fixed,
constant element of the Lie algebra. The first term in $\La$
contains the canonical 1-form for our theory and determines the
canonical brackets, while $f(n)$ describes the fluid dynamics. The
theory is invariant under gauge transformations with group
element $U$
\beq\label{eq317}
\begin{array}{rcl}
g &\to& U^{-1} g \\
A_\mu &\to& U^{-1}\left(A_\mu + \d_\mu\right) U\;.
\end{array}
\eeq
According to (\ref{eq314}), the current $J^\mu_a$ to which
$A_\mu^a$ couples is of the Eckart form (\ref{eq312})
\beq\label{eq318}
J^\mu_a = Q_a j^\mu
\eeq
with 
\beq\label{eq319}
Q\equiv Q_a T^a = g K g^{-1}\;.
\eeq
\par
For consistency of the gauge field dynamics $J^\mu\equiv
J_a^\mu T^a $ must be covariantly conserved. To satisfy the
general arguments of Section \ref{sec3a}, $j^\mu$ must be
divergenceless and then the Wong equation (\ref{eq39b}) will
follow. In Appendix \ref{app-a} we show that both conservation
laws are a consequence of invariance of the action with respect to
variations of the group element $g$; arbitrary variations of $g$
lead to covariant conservation of $ J^\mu$ (\ref{eq32}) while
the particular variation $\delta g= g K \lambda$ ensures that
$j^\mu$ is conserved (\ref{eq11}).  Therefore the Wong equation
(\ref{eq39b}) is also a consequence.
\par
Thus our model reproduces all the equations satisfied by the
current that were established in Section \ref{sec3a} from general
principles. Indeed the canonical (first) term of the Lagrangian
(\ref{eq314}) is like a Kirillov-Kostant $1$-form, which has
been previously used to give a Lagrangian for the point particle
Wong's equation (\ref{eq33}) \cite{ref-bal}. Moreover, as we show
in the Appendix \ref{app-b}, the canonical brackets implied by
the canonical 1-form ensure that the charge density algebra is
represented canonically
\beq\label{eq322b}
\{\rho_a(t,\vec{r}), \rho_b(t,\vec{r}') \} 
	= f_{abc}  \rho_c(t,\vec{r}) \delta(\vec{r}-\vec{r}')\;.
\eeq
\par
It remains to derive the Euler equation. This is accomplished by
varying $j^\mu$; stationary variation requires
\beq\label{eq323}
2 \tr \left[ Q (D_\mu g) g^{-1}\right] = \frac{u_\mu}{c^2} f'(n) 
\eeq
which we call the non-Abelian Bernoulli equation. The Euler
equation then follows, as in the Abelian case, by taking the curl
\beq\label{eq324}
\d_\mu \left(2\tr \left[ Q (D_\nu g) g^{-1}\right]\right) -
	\d_\nu \left(2\tr \left[ Q (D_\mu g) g^{-1}\right]\right) 
= \d_\mu \left(\frac{u_\nu}{c^2} f'(n)\right) 
	- \d_\nu \left(\frac{u_\mu}{c^2} f'(n)\right)\;.
\eeq
In the Appendix \ref{app-c} we show that manipulating the left
side allows rewriting (\ref{eq324}) as
\beq\label{eq324b}
2 \tr \left[ (D_\mu Q) (D_\nu g) g^{-1}\right] 
	+ 2 \tr \left[ Q F_{\mu\nu} \right] 
=  \d_\mu \left(\frac{u_\nu}{c^2} f'(n)\right)
      - \d_\nu \left(\frac{u_\mu}{c^2} f'(n)\right)\;.
\eeq
Finally, contracting with $j^\mu=n u^\mu$ and using (\ref{eq39b})
produces the relativistic, non-Abelian Euler equation
\beq\label{eq325}
	\frac{n u^\mu}{c^2}\d_\mu \left(u_\nu f'(n)\right) 
	- n \d_\nu  f'(n)=2 \tr\left[ J^\mu F_{\mu\nu}\right]\;.
\eeq
The left side is as in (\ref{eq212}) while the right side describes
the non-Abelian Lorentz force acting on the charged fluid.
\par
Apart from the gauge field contribution, the energy-momentum
tensor for the Lagrange density (\ref{eq314}) is the same as in
(\ref{eq213}) when (\ref{eq323}) is used to eliminate $\tr K g^{-1}
D_\mu g$
\beq\label{eq326}
T^{\mu\nu}=-g^{\mu\nu} \left(n f'(n) - f(n) \right) 
	+\frac{u^\mu u^\nu}{c^2} n f'(n)
\eeq
Just as in the Abelian case, the divergence of $T^{\mu\nu}$
entails two independent parts: one proportional to $u^\nu$ and
the other orthogonal to it
\beq\label{eq327}
\d_\mu T^{\mu\nu}= \d_\mu (n u^\mu) \frac{u^\nu f'(n)}{c^2}
	+ n\left[\frac{u^\mu}{c^2} \d_\mu (u^\nu f'(n)) 
	-\d^\nu f'(n) \right] \;.
\eeq
The first vanishes by the virtue of (\ref{eq11}) and the rest is
evaluated from Euler equation (\ref{eq325}), leaving
\beq\label{eq327b}
\d_\mu T^{\mu\nu}=  2 \tr \left[ J_\mu F^{\mu\nu}\right] 
\eeq
which is canceled by the divergence of the gauge-field
energy-momentum tensor
\beq
\d_\mu T^{\mu\nu}_{gauge}= - 2 \tr 
	\left[ J_\mu F^{\mu\nu}\right]\;.
\eeq
Thus energy-momentum conservation reproduces Abelian current
conservation (\ref{eq11}) and the non-Abelian Euler force equation
(\ref{eq326}), but the Wong equation (\ref{eq39b}) has to be
enforced additionally.  This is achieved by our Lagrangian
(\ref{eq314}).
\par
We record the nonrelativistic limit of (\ref{eq325});
using (\ref{eq-aux1}) -- (\ref{eq-aux2}), we find that the
nonrelativistic limit for the spatial component of (\ref{eq325})
gives the Euler equation with a non-Abelian Lorentz force
\beq
\d_t \vec{v} + \vec{v}\cdot\nabla \vec{v} = \vec{force}
	+ Q^a \vec{E}_a + \frac{\vec{v}}{c} \times Q^a \vec{B}_a 
\eeq
where $\vec{force}$ is the pressure force coming from the potential
$V$ (and is therefore Abelian in nature), while non-Abelian force
terms involve the chromoelectric and chromomagnetic fields
\beq
E^i_a = c F_{0i}^a \;, \qquad 
B^i_a = -\frac{c}{2} \epsilon_{ijk} F^a_{ij}
\eeq
\par
It is seen that our non-Abelian fluid moves effectively in a
single direction specified by $\vec{j}=\rho\vec{v}$. Nevertheless,
it experiences a non-Abelian Lorentz force. In the next subsection
we present a generalization wherein the non-Abelian fluid develops
several independent directions of motion.

%
%

\subsection{\label{sec3c}Generalization}

\hspace{0cm}\indent A generalization of our Lagrange density
(\ref{eq314}) that will give rise to several fluid components
carrying various densities and moving with various velocities
is obtained by choosing several directions in the Lie algebra,
$K_{(s)}$, and coupling to different Abelian currents
\cite{ref-jackiw-2}
\beq\label{eq328}
\begin{array}{rcl}
\La&=& \displaystyle\sum\limits_{s=1}^{r} j^\mu_{(s)} 
	2 \tr \left[ K_{(s)} g^{-1}D_\mu g\right]
	-f(n_{(1)},n_{(2)},\ldots, n_{(r)}) + \La_{gauge} \\
j^\mu_{(s)}&=& (c\rho_{(s)} , \rho_{(s)} \vec{v}_{(s)} )
	=  n_{(s)} u^\mu_{(s)}  \\
	u^\mu_{(s)} u_{(s) \mu} &=& c^2  \\
n_{(s)}&=& \sqrt{j^\mu_{(s)} j_{(s) \mu}/c^2 }
\end{array}
\eeq
(Sums over $s$ are indicated explicitly; the summation convention
does not apply.) Evidently the current which couples to the gauge
potential is now
\beq\label{eq329}
J^\mu=\sum\limits_{s=1}^{r} Q_{(s)} j^\mu_{(s)} \;,
\qquad \mbox{with}\qquad
Q_{(s)} = g K_{(s)} g^{-1}\;.
\eeq
Arbitrary variation of $g$ ensures that (\ref{eq329}) is
covariantly conserved, but we also need the conservation of
individual $j^\mu_{(s)}$.  This is achieved by choosing special
forms for $\delta_{(s)} g = g K_{(s)} \lambda_{(s)}$ which will
work provided that different $K_{(s)}$ commute (see Appendix
\ref{app-a}). Therefore, we choose the $K_{(s)}$ to belong to
the Cartan subalgebra of the Lie algebra and the total number of
different channels equals the rank $r$ of the group. With this
choice for the $K_{(s)}$, special variations of $g$ ensure
\beq\label{eq329b}
\d_\mu j^\mu_{(s)} = \d_t \rho_{(s)} + \nabla\cdot (\rho_{(s)}
\vec{v}_{(s)} )=0 
\eeq
The Wong equation which follows from the conservation of the
non-Abelian current now reads
\beq\label{eq330}
\sum\limits_{s=1}^{r} j^\mu_{(s)}D_\mu Q_{(s)}=0 \;.
\eeq
Varying the individual $j^\mu_{(s)}$ produces the Bernoulli
equations
\beq\label{eq331}
2 \tr \left[ Q_{(s)} (D_\mu g) g^{-1}\right] = \frac{u_\mu}{c^2}
f^{(s)}
\qquad \mbox{where } \qquad f^{(s)} \equiv 
	\frac{\d}{\d n_{(s)} } f(n_{(1)},n_{(2)},\ldots,n_{(r)})
\eeq
Again, the curl of the above can be cast in the form 
\beq\label{eq332}
\frac{1}{c^2} \left\{ \d^\mu \left(u^\nu_{(s)} f^{(s)}\right)
      - \d^\nu \left(u^\mu_{(s)} f^{(s)}\right)\right\}
=2 \tr \left[ (D^\mu Q_{(s)}) (D^\nu g) g^{-1}\right]
      + 2 \tr \left[ Q_{(s)} F^{\mu\nu} \right]
\;.
\eeq
When contracted with $j^\mu_{(s)}=n_{(s)} u^\mu_{(s)}$, this leaves
\beq\label{eq333}
\frac{n_{(s)}u^{\mu}_{(s)} }{c^2} 
	\d_\mu \left(u^\nu_{(s)} f^{(s)}\right)
      - n_{(s)} \d^\nu  f^{(s)}
=2 \tr \left[j^{\mu}_{(s)} (D_\mu Q_{(s)}) (D^\nu g) g^{-1}\right]
      + 2 \tr \left[j_{\mu (s)} Q_{(s)} F^{\mu\nu} \right]
\;.
\eeq
However, unlike in the single channel case, the right side does
not simplify since $j_{\mu (s)} Q_{(s)} $ cannot be replaced by
$J_\mu$ because the latter requires summing over $s$. Also the
first right-hand term in (\ref{eq333}) does not vanish since
(\ref{eq330}) requires summation over $s$.
\par
The energy-momentum tensor is 
\beq\label{eq336}
T^{\mu\nu}=-g^{\mu\nu} \left(
	\sum\limits_{s=1}^{r} n_{(s)} f^{(s)} - f\right) 
	+ \sum\limits_{s=1}^{r} 
	\frac{ u^\mu_{(s)} u^\nu_{(s)} }{c^2} n_{(s)} f^{(s)}\;.
\eeq
Its divergence of course reproduces (\ref{eq327})
\beq\label{eq337}
\d_\mu T^{\mu\nu} =  \sum\limits_{s=1}^{r}\left\{ 
	\left(\d_\mu (n_{(s)}u^\mu_{(s)})\right) 
	\frac{u^\nu_{(s)} f^{(s)}}{c^2} 
	+  n_{(s)}\left[ \frac{u^\mu_{(s)} }{c^2}
	\d_\mu\left( u^\nu_{(s)} f^{(s)}\right) 
	-  \d^\nu  f^{(s)}\right]
\right\}\;.
\eeq
The first term in the brackets vanishes according to (\ref{eq329b})
and the remainder is evaluated from (\ref{eq333}) as
\begin{displaymath}
\sum\limits_{s=1}^{r}\left( 2 \tr \left[
	(j^{\mu}_{(s)} D_\mu Q_{(s)}) (D^\nu g) g^{-1}\right] 
	+ 2 \tr \left[j_{\mu (s)} Q_{(s)} F^{\mu\nu} \right]
\right)
\end{displaymath}
Since now we are summing over all channels, it follows from
(\ref{eq329}) and (\ref{eq330}) that, as before,
\beq
\d_\mu T^{\mu\nu} =   2 \tr\left[ J_\mu F^{\mu\nu} \right]
\eeq
\par
A more transparent picture of what is happening is given if the
dynamical potential separates
\bea\label{eq334}
f(n_{(1)},\ldots,n_{(r)}) &=& \sum\limits_{s=1}^{r} f_{(s)}(n_{(s)})\\
f^{(s)}&=& f'_{(s)}
\eea 
Then the left side of (\ref{eq333}) refers only to variables
labeled $s$, while the right side may be rewritten with the help of
(\ref{eq328}) and (\ref{eq330}) to give
\beq\label{eq335}
\frac{n_{(s)} u^{\mu}_{(s)}}{c^2} \d_\mu \left(
	u^\nu_{(s)} f'_{(s)}\right) - n_{(s)} \d^\nu  f'_{(s)}
=2 \tr \left[J_{\mu} F^{\mu\nu} \right]
- 2 \sum\limits_{s'\neq s}^{r} \tr \left[j_{\mu (s')}
      \left( Q_{(s')} F^{\mu\nu} + (D^\mu Q_{(s')}) (D^\nu g)
g^{-1}\right)\right] \;.
\eeq
Thus in the addition to the Lorentz force, there are forces
arising from the other channels $s'\neq s$.
Note that for separated dynamics (\ref{eq334}), the energy-momentum
tensor also separates
\beq\label{eq339}
T^{\mu\nu}=\sum\limits_{s=1}^{r} T^{\mu\nu}_{(s)}
=\sum\limits_{s=1}^{r}\left\{ -g^{\mu\nu}
	\left( n_{(s)} f'_{(s)} - f_{(s)}(n_{(s)}) \right)
	+  \frac{ u^\mu_{(s)} u^\nu_{(s)} }{c^2} 
	n_{(s)} f'_{(s)}\right\}\;.
\eeq
but the divergence of individual $T^{\mu\nu}_{(s)}$ does not
vanish. This reflects that energy is exchanged between the
different channels and with the gauge field, as is also evident
from the equation of motion (\ref{eq335}). It is clear that this
fluid moves with $r$ different velocities $\vec{v}_{(s)}$.
\par
The single-channel Euler equation (\ref{eq325}) is expressed in
terms of physically relevant quantities (currents, chromomagnetic
fields); the many-channel equation (\ref{eq333}) involves,
additionally, the the gauge group element $g$.  One may simplify
that equation by going to special gauge, for example $g=I$,
so that the right side of (\ref{eq333}) reduces to
\beq\label{eq340}
2 \tr \left[j_{\mu (s)} (D^\mu Q_{(s)}) (D_\nu g) g^{-1}\right]
      + 2 \tr \left[j^{\mu}_{(s)} Q_{(s)} F_{\mu\nu} \right]
= 2 \tr \left[ K_{(s)} j^\mu_{(s)} (\d_\mu A_\nu-\d_\nu A_\mu)\right]
\eeq
while the Wong equation (\ref{eq330}) becomes 
\beq\label{eq341}
\sum\limits_{s=1}^{r} j^\mu_{(s)} [A_\mu,K_{(s)}]=0
\eeq
It is interesting that in this gauge the non-linear terms in
$F^{\mu\nu}$ disappear.

%
%

\section{\label{sec4}Non-Abelian Fluids with a Field Substratum}

\hspace{0cm}\indent In the Abelian case, the M\"adelung
parametrization (\ref{eq28}) of the Schr\"odinger equation gives
the conventional nonrelativistic Euler equation, [even while the
forces are derived from a potential that depends on density and
(unconventionally) on its derivatives] (see Section \ref{sec2}). We
are therefore led to examine a M\"adelung-like construction for
a non-Abelian, ``colored'' Schr\"odinger equation
\beq\label{eq41}
i\hbar\: \d_t \psi = -\frac{\hbar^2}{2m} \nabla^2 \psi
\eeq
We consider only the free, nonrelativistic case, and the
``non-Abelian'' structure resides solely in the fact that the
$\psi$ is a multi-component object, transforming under some
representation of a group. The color degrees of freedom also lead
to the conserved color current
\beq\label{eq42}
J^\mu_a=(c \rho_a, \vec{J}_a)\;,\qquad \rho_a
	= i \psi^\dagger T^a \psi\;, \qquad
\vec{J}_a = \frac{\hbar}{m} \mathrm{Re}\,\psi^\dagger T^a 
	\nabla \psi\;.
\eeq
Of course, the singlet current 
\beq\label{eq43}
j^\mu=(c \rho, \vec{j})\;,\qquad \rho
	= \psi^\dagger \psi\;, \qquad
\vec{j} = \frac{\hbar}{m}\mathrm{Im}\, 
	\psi^\dagger \nabla \psi\;.
\eeq
is also conserved. For definiteness and simplicity, we shall henceforth
assume that the group is $SU(2)$ and that the representation is the
fundamental one: $T^a={\sigma^a}/{(2i)}$, $\{T^a,T^b\}=-\delta^{ab}/2$.
We shall also set the mass $m$ and Planck's constant $\hbar$ to unity. The
non-Abelian analog of the M\"adelung decomposition
(\ref{eq28}) is
\beq\label{eq44} 
\psi=\sqrt{\rho} g u
\eeq
where $\rho$ is the scalar $\psi^\dagger\psi$, $g$ is a
group element, and $u$ is a constant vector that points in
a fixed direction [e.g.,  for $SU(2)$ $u$ could be $
\renewcommand\baselinestretch{1} \bmx{1} 1 \\ 0 \emx$, then $i
u^\dagger T^a u= \delta^{a3}/2$].  The singlet density is $\rho$,
while the singlet current $\vec{j}$ is
\beq\label{eq45}\label{eq46}
\vec{j}=\rho\vec{v}\;, \qquad \vec{v}\equiv 
	-i u^\dagger  \:g^{-1}\nabla g\:u
\;.
\eeq
With the decomposition (\ref{eq44}), the color density (\ref{eq42})
becomes
\beq\label{eq47}\label{eq48}
\rho_a=Q_a \rho\;,\qquad Q_a= i u^\dagger\: g^{-1}T^a g \: u = i R_{ab}
u^\dagger T^b u =\ R_{ab} t^b/2
\eeq
where $R_{ab}$ is in the adjoint representation of the group and
the unit vector $t^a$ is defined as $t^a/2= i u^\dagger T^a u$. On
the other hand, the color current reads
\beq\label{eq49}
\vec{J}_a =\frac{1}{2} \rho R_{ab} u^\dagger \:(T^b 
	\: g^{-1}\nabla g + g^{-1}\nabla g \: T^b) \: u \;,
\eeq
which with the introduction of
\bea\label{eq49b}
g^{-1}\nabla g &\equiv& -2\vec{v}^a T^a\\
\vec{v}&=&\vec{v}^a t^a
\eea 
becomes
\beq\label{eq410}
\vec{J_a}=\frac{\rho}{2} R_{ab} \vec{v}^b\;.
\eeq
Unlike the Abelian model, the vorticity is nonvanishing
\beq\label{eq410a}
\nabla\times\vec{v}^a = \epsilon^{abc} \vec{v}^b\times \vec{v}^c
\eeq
\par
A difference between the M\"adelung approach and the previous
particle based one is that the color current is not proportional
to the singlet current. Equation (\ref{eq410}) may be written as
\beq
\vec{J}_a =  Q_{a}\rho \vec{v} 
	+\frac{\rho}{2} R_{ab} \vec{v}^b_\perp
\eeq
where the ``orthogonal'' velocity $\vec{v}_\perp^a$ is defined as
\beq
\vec{v}^a_\perp=\left(\delta^{ab}-t^a t^b\right) \vec{v}^b\;.
\eeq
Equation (\ref{eq410a}) shows that color current possesses
components that are orthogonal to the singlet current.
\par
In Appendix \ref{app-d} we derive for the $SU(2)$ case
the decomposition of the Schr\"odinger equation with the
parametrization (\ref{eq44}). Two equations emerge: one regains
the conservation of the Abelian current (\ref{eq11}) and the
other is the ``Bernoulli'' equation
\beq\label{eq413}
\left(  g^{-1} \d_t g\right)^a = \left[ \vec{v}^b\cdot \vec{v}^b 
	- \frac{\nabla^2 \sqrt{\rho}}{\sqrt{\rho}}\right] t^a 
	+ \frac{1}{\rho} \nabla\cdot \left(\rho \epsilon^{abc} 
	\vec{v}^b t^{c}\right) 
\eeq
It is further verified that the covariant conservation
of the color current is enforced by both (\ref{eq11}) and
(\ref{eq413}). However, there is no Wong equation because the
color current is not proportional to the conserved singlet
current. Finally, using the identity, which is a consequence of
the definition (\ref{eq49b})
\beq\label{eq414}
\d_t \vec{v}^a = -\frac{1}{2}\nabla \left( g^{-1} \d_t g
\right)^a + \epsilon^{abc} \vec{v}^b \left(g^{-1} \d_t g
\right)^c 
\eeq
one can deduce an Euler equation for $\d_t\vec{v}^a$ from
(\ref{eq413}).
\par
We record the energy and momentum density
\bea
\label{eq415}
\mathcal{E}&=&\frac{1}{2}\nabla{\psi^\dagger} \cdot \nabla \psi 
= \frac{1}{2}\rho \vec{v}^a\cdot \vec{v}^a 
	+ \frac{\nabla\rho \cdot \nabla\rho}{8 \rho}
\\
\label{eq416}
\vec{\mathcal{P}}&=&\frac{i}{2}\left(
	\nabla \psi^\dagger \psi
	-\psi^\dagger \nabla \psi \right)= \rho \vec{v}
\eea
Both parallel and orthogonal components of the velocity contribute
to the energy density but only the parallel component $\vec{v}$
contributes to the momentum density. It is clear that within
the present approach the fluid color flows in every direction in
the group space, but the mass density is carried by the unique
velocity $\vec{v}$.  This is in contrast to our previous approach
where all motion is in a single direction (Section \ref{sec3b})
or at most in the directions of the Cartan elements of the Lie
algebra (Section \ref{sec3c}).
\par
The difference between the two approaches is best seen from a
comparison of Lagrangians. For the color Schr\"odinger theory in
the M\"adelung representation
\bea\label{eq417}
\La_{Schrodinger}&=&\frac{i}{2}\left(\psi^\dagger\d_t\psi 
	- \d_t \psi^\dagger \:\psi \right) 
	-\frac{1}{2} \nabla\psi^\dagger\cdot \nabla \psi 
\\
\label{eq418}
&=& i\rho \:u^\dagger \:g^{-1}\d_t g\: u
	-\frac{1}{2}\rho\vec{v}^a\cdot \vec{v}^a 
	-\frac{(\nabla\rho)^2}{8 \rho}
\eea
With $u\otimes u^\dagger \equiv I/2 - 2 i K$, the free part of
the above reads
\beq\label{eq419}
\La^0_{Schrodinger} = \rho \: 2\tr \left[ K g^{-1}\d_t g\right]
-\frac{1}{2} \rho \vec{v}^a\cdot \vec{v}^a
\eeq
On the other hand, the free part of the Lagrange density
(\ref{eq314}) in the nonrelativistic limit is
\bea
\La^0&=&\rho \:2\tr\left[ K g^{-1}\d_t g \right] 
	+ \rho\vec{v}\cdot 2 \tr\left[ K g^{-1}\nabla g\right] 
	- \sqrt{\rho^2 (c^2-v^2)}
\nonumber \\
&\approx& \rho \:2\tr\left[ K g^{-1}\d_t g \right] 
	+ \rho\vec{v}\cdot 2 \tr\left[ K g^{-1}\nabla g\right] 
	- \rho c^2+\frac{1}{2}\rho v^2
\nonumber \\
\label{eq420}
&=& \rho \:2\tr\left[ K g^{-1}\d_t g \right] 
	-\frac{1}{2}\rho v^2 - \rho c^2
\eea
where we have used $\vec{v}=-2 \tr\left[ K g^{-1}\nabla g\right]$,
which follows upon the variation of $\vec{v}$, in the next-to-last
equality above. Thus the canonical $1$-form is the same for both
models while the difference resides in the velocity dependence
of their respective Hamiltonians.  Only the singlet $\vec{v}$
enters (\ref{eq420}) while the M\"adelung construction uses the
group vector $\vec{v}^a$.
\par
Finally, note that while the Euler equation, which emerges when
(\ref{eq413}) and (\ref{eq414}) are combined, intricately couples
all directions of the fluid velocity $\vec{v}^a$, it does admit
the simple solution $\vec{v}^a=\vec{v}t^a$, with $\vec{v}$ obeying
the Abelian equations that arise from (\ref{eq27})-(\ref{eq28b}).

\section{\label{sec5}Discussion}

In this paper, we have presented in Section \ref{sec3} two distinct
non-Abelian generalizations of ordinary particle based Abelian
fluid dynamics.  Both versions use a fluid generalization of
the Kirillov-Kostant form that naturally encodes the algebra of
charge densities (\ref{eq322b}), which is needed if the charge
density is to be identified as the generator of non-Abelian
symmetry transformations.  In Section \ref{sec3c} we generalized
the first version, given in Section \ref{sec3b} so that the density
is specified in terms of a set of Abelian densities equal in number
to the rank $r$ of the Lie algebra, rather than a single density as
in the first version.  Since the charge density at a point in the
fluid is an element of the Lie algebra, diagonalization shows that
an invariant specification must use $r$ eigenvalues. Alternatively,
we may use the $r$ Casimir invariants of $\rho_a$ at a point to
characterize it. Therefore the appearance of $r$ Abelian currents
in the Lagrangian is entirely natural. This is also in accord with
what happens with the Kirillov-Kostant form where our $j^\mu_{(s)}$
are replaced by the fixed weights of a representation of the Lie
algebra and lead to that representation upon quantization.
\par
The two versions also differ in the formula for
the current.  Our first version, which is mathematically more
concise, gives the non-Abelian Eckart decomposition, (\ref{eq318}),
while the second version does not allow the
factorization of the current into a non-Abelian charge density
and an Abelian velocity, rather it is a sum of such factorized
expressions -- a generalized Eckart decomposition as in (\ref{eq319}). The
Eckart decomposition shows that we can choose a
local Lorentz frame for which a given fluid element can be brought
to rest. The charge density is then related to the charge carried
by this element. By
contrast, in the second version, with a generalized Eckart decomposition, we see that even if we choose
a frame where one Lie algebra component of the velocity is zero,
the other color velocities need not be. Thus the latter applies
to a situation where color separating flows can occur. Physically,
it is not yet clear what kinematic regimes of a quark-gluon plasma,
for example, would admit or require such flows.

\par
We note that the currents we have obtained are the non-Abelian
analogues of the irrotational part of the Abelian current, even
though the vorticity is nonvanishing. The other components can be
easily incorporated, if needed , by generalizing the Lagrangian in
(\ref{eq328}) as
\begin{equation}
\La = \sum\limits_{s=1}^{r} j^\mu_{(s)} \left\{
2 \tr \left[ K_{(s)}
g^{-1}D_\mu g\right] +a_{\mu (s)}\right\}
    -f(n_{(1)},n_{(2)},\ldots, n_{(r)}) + \La_{gauge}
\label{eq511}
\end{equation}
where $a_{\mu (s)}$ is given by
\beq
a_{\mu (s)} = \alpha_{(s)}\partial_\mu \beta_{(s)}\label{eq512}
\eeq
The final fluid equations remain unchanged.
\par
We have also derived in Section \ref{sec4} a field based
fluid mechanics by extending the M\"adelung construction to
the non-Abelian situation.  Here the ``Euler'' equations are
much less appealing because they involve velocities in all
group directions. We know of no compelling physical reason for
preferring this field based model over the particle based one, even
though in the Abelian case it coincides with the particle-based
construction.

\section*{Acknowledgements}
This work was supported in part by DOE grants Nos. 
DE-FC02-94-ER40818, DE-FG02-91-ER40676, by 
NSF grant number PHY-0070883
and by a PSC-CUNY grant.

%
%

\begin{appendix}
\section{\label{app-a}Variations of $g$}
We determine the variation of 
\beq\label{eqa1}
I^0=\int\dd t \dd r \sum\limits_{s=1}^{r} j^\mu_{(s)}\,
	2\tr K_{(s)} g^{-1}D_\mu g 
\eeq
when $g$ is varied either arbitrarily or in the specific manner
\beq\label{eqa2}
g^{-1}\delta g = K_{(s')} \lambda
\eeq
where $\lambda$ is an arbitrary function on space-time. This
will provide the needed results (\ref{eq11}) and (\ref{eq32})
for the single channel situation as well as (\ref{eq329b}) and
(\ref{eq330}) for many channels.
\par
Recall the definitions $Q_{(s)}=g K_{(s)}g^{-1}$ and 
$D_\mu g = \d_\mu g + A_\mu g$, which implies
$D_\mu g^{-1} = \d_\mu g^{-1} - g^{-1} A_\mu$. 
First, the variation of $g^{-1}D_\mu g$ is established
\beq\label{eqa3}
\delta\left(g^{-1}D_\mu g\right) 
	= -g^{-1}\delta g g^{-1} D_\mu g 
	+ g^{-1}D_\mu \delta g
\eeq
To evaluate the last term, note that $D_\mu \delta g =
D_\mu\left(g g^{-1} \delta g\right)= (D_\mu g) g^{-1} \delta g +
g D_\mu\left(g^{-1} \delta g\right)$.
Thus
\beq\label{eqa4}
\delta\left(g^{-1}D_\mu g\right) = \d_\mu \left(g^{-1} \delta g\right) + 
	[ g^{-1} D_\mu g,g^{-1} \delta g]\;.
\eeq
Inserting (\ref{eqa4}) into the variation of $I^0$ in (\ref{eqa1}),
integrating by parts, and rearranging the trace with $K_{(s)}$ gives
\beq\label{eqa5}
\delta I^0 = -\int \dd t \dd r \sum\limits_{s=1}^{r} \left(\d_\mu
j^\mu_{(s)} \, 2 \tr K_{(s)} g^{-1} \delta g + j^\mu_{(s)} 2 \tr \left[
g^{-1} D_\mu g , K_{(s)}\right] g^{-1}\delta g\right)\;.
\eeq
\par
Considering first arbitrary variations: the vanishing of $\delta
I^0$ requires
\beq\label{eqa6}
\sum\limits_{s=1}^{r}\left(\d_\mu j^\mu_{(s)} K_{(s)} 
	+ j^\mu_{(s)} \left[ g^{-1} D_\mu g , K_{(s)}\right]\right)=0
\eeq
or, after sandwiching the above between $g\ldots g^{-1}$,
\beq\label{eqa7}
\sum\limits_{s=1}^{r}\left(\d_\mu j^\mu_{(s)} Q_{(s)} 
	+ j^\mu_{(s)} \left[ \left(D_\mu g\right) 
	g^{-1}, Q_{(s)}\right]\right)=0\;.
\eeq
Finally we verify that 
\beq\label{eqa8}
\left[ D_\mu g g^{-1}, Q_{(s)}\right]= D_\mu Q_{(s)}\;,
\eeq
so that the desired results (\ref{eq32})  and (\ref{eq330}) follow
\beq\label{eqa9}
\sum\limits_{s=1}^{r}\left(\d_\mu j^\mu_{(s)} Q_{(s)} 
	+ j^\mu_{(s)}  D_\mu Q_{(s)} \right)
	=D_\mu\left(\sum\limits_{s=1}^{r} j^\mu_{(s)}
	Q_{(s)}\right)= D_\mu J^\mu =0  \;.
\eeq  
\par
Next we consider the specific variation (\ref{eqa2}) and separate
the sum (\ref{eqa5}) into the term $s=s'$ and $s\neq s'$. After
a rearrangement of the last term in (\ref{eqa5}), we get
\bea\label{eqa10}
\delta I^0 &=& -\int\dd t \dd r \left( \d_\mu j^\mu_{(s')} \, 2\tr
K_{(s')}K_{(s')}\lambda +  j^\mu_{(s')} \, 2\tr g^{-1}D_\mu g
[K_{(s')},K_{(s')}]\lambda\right)
\nonumber \\ && 
+\sum\limits_{s\neq s'} \left( \d_\mu j^\mu_{(s)} \, 2\tr 
K_{(s)}K_{(s')}\lambda +  j^\mu_{(s)} \, 2\tr g^{-1}D_\mu g 
[K_{(s)},K_{(s')}]\lambda\right)
\eea
The first commutator vanishes; so does the second when $K_{(s)}$
and $K_{(s')}$ commute, i.e., when they belong to the
Cartan subalgebra. Also $2\tr K_{(s)}K_{(s')}= - K_{(s)}^a
K_{(s')}^a $; for $s'=s$ this is constant, while for $s'\neq s$
it vanishes when it is arranged that distinct elements of the
Cartan algebra are selected.  Thus for stationary variations
$j^\mu_{(s)}$ must be conserved, and (\ref{eq11}) as well as
(\ref{eq32}) are validated.

%
%
\section{\label{app-b}Charge density algebra}

The portion of the Lagrange density (\ref{eq314}) that determines
the Poisson bracket is
\beq\label{eqb1}
\La_{canonical}=\rho \,2\tr K g^{-1}\d_t g = \rho \,2\tr Q \d_t g g^{-1}\;.
\eeq
With a parametrization of the group element, e.g., 
$g(\varphi) = e^{T^a \varphi_a}$, one sees that $\left(\d_t
g\right) g^{-1}$ has the form $-\d_t \varphi_a C^{a}_{\;
b}(\varphi) T^b$, where the non-singular matrix $C^{a}_{\;
b}(\varphi)$ is defined by
\beq\label{eqb2}
C^{a}_{\; b}(\varphi) T^b = -\frac{\d g(\varphi)}{\d \varphi_a}
g^{-1}(\varphi)\;.
\eeq
Thus 
\beq\label{eqb3}
\La_{canonical}=\rho \d_t \varphi_a C^{a}_{\; b}Q_b 
	= \d_t \varphi_a C^{a}_{\; b} \rho_b
\eeq
and the momentum conjugate to $\varphi_a$ is
\beq\label{eqb3b}
\Pi^a= C^{a}_{\; b} \rho_b\;.
\eeq
With inverse to $C^{a}_{\; b}$ defined as $c^{b}_{\; a}\:$, 
\beq\label{eqb4}
\rho_a= c^{a}_{\; b} \Pi^b\;.
\eeq
\par
The non-Abelian charge density $\rho_a$ is a function of
$(t,\vec{r})$ and for (\ref{eq322b}) we need the bracket with
another density evaluated at $(t,\vec{r}')$. Since the dependence
of $ c^{a}_{\; b}$ on $\varphi$ involves no spatial derivatives
of $\varphi$, it is clear that the brackets will be local in
$\vec{r}-\vec{r}'$; just as is the bracket between $\varphi$
and $\Pi$.
\bea\label{eqb5}
\{\rho_a(t,\vec{r}),\rho_b(t,\vec{r}')\}&=& \left( 
	c^{b}_{\; b'}\frac{\d  c^{a}_{\; a'}}{\d \varphi_{b'}} \Pi^{a'}
	- \;a \leftrightarrow b\right)\delta \left(\vec{r}-\vec{r}'\right)
\nonumber\\
&=& \left(
      -c^{b}_{\; b'}c^{a}_{\; c'}\frac{\d  C^{c'}_{\; c''}}
	{\d \varphi_{b'}}c^{c''}_{\; a'} \Pi^{a'}
      - \;a \leftrightarrow b\right)\delta \left(\vec{r}-\vec{r}'\right)
\nonumber\\                                                                                                 
&=& \left(-c^{b}_{\; b'}c^{a}_{\; c'}\frac{\d  C^{c'}_{\; c''}}{\d
\varphi_{b'}}\rho_{c''}
      - \;a \leftrightarrow b\right)\delta \left(\vec{r}-\vec{r}'\right)
\eea
To evaluate the derivative with respect to $\varphi$, return to
(\ref{eqb2}) and observe
\bea\label{eqb6}
\frac{\d  C^{c'}_{\; c''}}{\d \varphi_{b'}} &=& \frac{\d}{\d \varphi_{b'}}
	\left(2\tr\frac{\d g}{\d \varphi_{c'}} g^{-1} T^{c''} \right)
\nonumber \\
&=& 2 \tr\left( \frac{\d^2 g}{\d \varphi_{c'}
	\d \varphi_{c'}} g^{-1} - C^{c'}_{\; d'} T^{d'}  
	C^{b'}_{\; d''} T^{d''}\right) T^{c''}
\eea
The first term in the parentheses is symmetric in $(b',c')$; when
inserted in (\ref{eqb5}) it produces a symmetric contribution in
$(a,b)$ and does not contribute when antisymmetrization in $(a,b)$
is effected. What is left establishes (\ref{eq314}).
\bea\label{eqb7}
\{\rho_a(t,\vec{r}),\rho_b(t,\vec{r}')\}&=& 
\left(
      c^{b}_{\; b'} c^{a}_{\; c'} C^{c'}_{\; d'} C^{b'}_{\; d''}
	2 \tr T^{d'} T^{d''} T^{c''} \rho_{c''} 
      - \;a \leftrightarrow b\right)
	\delta \left(\vec{r}-\vec{r}'\right)
\nonumber\\
&=& 
-\left( 2\tr T^{a}T^{b}T^{c''}\rho_{c''} - \;a \leftrightarrow
b\right)\delta \left(\vec{r}-\vec{r}'\right)
\nonumber\\
&=& - 2 \tr f^{abd} T^{d} T^{c''}\rho_{c''} 
	\delta \left(\vec{r}-\vec{r}'\right)
\nonumber\\
&=&  f^{abc} \rho_{c}(t,\vec{r})
	\delta \left(\vec{r}-\vec{r}'\right)
\eea

%
%

\section{\label{app-c} Manipulating equation (\ref{eq324})}
Observe that the first term in  (\ref{eq324}) equals
\beq\label{eqc1}
\d_\mu 2 \tr \left[ Q (D_\nu g) g^{-1}\right]=
2 \tr \left[ (D_\mu Q) (D_\nu g) g^{-1}+ 
Q (D_\mu D_\nu g ) g^{-1} - Q  (D_\nu g) g^{-1}  (D_\mu g) g^{-1}\;.
\right]
\eeq
The first term on the right side is rewritten with the help of
(\ref{eqa8}) and combined with the last term, leaving
\begin{displaymath}
2\tr \left( Q (D_\mu D_\nu g) g^{-1} - Q (D_\mu g) g^{-1} (D_\nu g) g^{-1}
\right)
\end{displaymath}
After antisymmetrization in $(\mu,\nu)$, the left side of
(\ref{eq324}) reads
\beq\label{eqc2}
2\tr Q \left(([D_\mu, D_\nu]g) g^{-1} 
	-[ (D_\mu g) g^{-1} ,(D_\nu g) g^{-1}]\right)=
	2\tr\left( Q F_{\mu\nu} 
	- [Q,(D_\mu g) g^{-1}] (D_\nu g) g^{-1}\right)
\eeq
When (\ref{eqa8}) is used again, (\ref{eqc2}) becomes the left
side of (\ref{eq325}).

%
%

\section{\label{app-d} Non-Abelian M\"adelung Parametrization}

When (\ref{eq44}) is inserted into (\ref{eq41}), and use is made
of the definition (\ref{eq49b}), we find in the $SU(2)$ case
\beq\label{eqd1}
\frac{1}{2}i\d_t \rho \,u + i\rho (g^{-1}\d_t g)^a T^a \,u = 
	-\frac{1}{2}\sqrt{\rho} \nabla^2\sqrt\rho \,u 
	+\nabla (\rho \vec{v}^a) T^a \,u 
	+\frac{1}{2}\rho \vec{v}^a\cdot\vec{v}^a \, u 
\eeq
Next (\ref{eqd1}) is premultiplied by $u^\dagger$, where it implies
\beq\label{eqd2}
i\d_t \rho  + \rho (g^{-1}\d_t g)^a {t^a}  =
	-\sqrt{\rho} \nabla^2\sqrt\rho 
	-i \nabla \left(\rho \vec{v}^a {t^a}\right)
 	+\rho \vec{v}^a\cdot\vec{v}^a \;.
\eeq
The imaginary part reproduces the continuity equation for the
singlet current (\ref{eq45}), while the real part gives
\beq\label{eqd3}
 (g^{-1}\d_t g)^a t^a  =
-\frac{1}{\sqrt{\rho}} \nabla^2\sqrt\rho  
 +\vec{v}^a\cdot\vec{v}^a \;.
\eeq
To obtain further information, we premultiply (\ref{eqd1}) with
$u^\dagger T^{b}$. This gives
\beq\label{eqd4}
\d_t \rho t^b - i\rho (g^{-1}\d_t g)^b  
	+\rho (g^{-1}\d_t g)^a \epsilon^{bac} {t^c} 
= i\sqrt{\rho} \nabla^2\sqrt\rho t^b 
	- \nabla \cdot\left(\rho \vec{v}^b \right)
	-i \nabla\cdot\left( \rho \vec{v}^a\right) \epsilon^{bac} {t^c}
 	-i\rho \vec{v}^a\cdot\vec{v}^a t^b\;.
\eeq
The imaginary part gives (\ref{eq413}) while the real part is
identically satisfied by virtue of (\ref{eq11}) and (\ref{eq413}).

\end{appendix}


\begin{thebibliography}{99}
\bibitem{ref-6}
When comparison is made with the usual expression for the
energy-momentum tensor of a relativistic fluid, we see that $n
f'(n)-f(n)$ is the pressure and $n f'(n)$ is the pressure summed
with the energy density, i.e., $f(n)$ is the energy
density; see Ref.  \cite{ref-jackiw-1}.
\bibitem{ref-jackiw-2}
In a previous paper, we presented a similar candidate Lagrangian
for non-Abelian fluids, except that the $K_{(s)}$ comprised
all generators for a subgroup $H$ of the group $G$ to which $g$
belongs, with the additional requirement that $G/H$ be a symmetric
space. This construction is based on a non-Abelian version of
the Clebsch parametrization. But the equations that follow do
not have a clear physical interpretation, so we have not pursued
this approach.
R. Jackiw, V. P. Nair and S.-Y. Pi, \textit{Phys. Rev. D}
\textbf{62}, 085018 (2000).
\end{thebibliography}
\end{document}